\documentclass[reprint, amsmath,amssymb,pra]{revtex4-1}

\usepackage{graphicx}
\usepackage{dcolumn}
\usepackage{bm}

\begin{document}

\preprint{APS/123-QED}

\title{Coherent population oscillation produced by saturating probe and pump\\ fields on the intercombination Line}

\author{A. Vafafard$^{1,2}$}
\altaffiliation{azar.vafafard@okstate.edu}

\author{M. Mahmoudi$^{2}$}
\author{G. S. Agarwal$^{1}$}
\affiliation{$^{1}$Department of Physics, Oklahoma State University, Stillwater, Oklahoma
74078, USA}
\affiliation{$^{2}$Department of Physics, University of Zanjan, University Blvd, 45371-38791, Zanjan, Iran}
\date{\today }

\begin{abstract}
We present a theoretical study of the experiments on coherent population oscillations and coherent population trapping on the intercombination line of $^{174}Yb$. The transition involves a change of the spin and thus can not be interpreted in terms of an effective Lambda system. The reported experiments are done in the regime where both pump and probe fields can saturate the transition. We demonstrate by both numerical and analytical calculations the appearance of the interference minimum as both pump and probe start saturating the transition. We present an analytical result for the threshold probe power when the interference minimum can appear. We also present detailed study of the appearance of the interference minimum when magnetic fields are applied. The magnetic fields not only create Zeeman splittings but in addition make the system open because of the couplings to other levels. We show the possibility of interference minimum at the position of subharmonic resonances. 
\begin{description} 
\item[PACS numbers]
42.50.Gy, 42.50.Hz, 42.50.Nn
\end{description}
\end{abstract}


\maketitle

\section{\label{sec:level1}Introduction}

Atomic coherence effects, induced by laser fields, such as the coherent population trapping (CPT) \cite{Arimondo}, the coherent population oscillation (CPO)\cite{BoydCPO}, the electromagnetically induced transparency (EIT)\cite{HarrisEIT}, have become increasingly popular because of their very wide applications in lasing without inversion\cite{AgarwalLWI}, enhancement of refractive index \cite{ScullyRef}, slow light\cite{Slow, SlowA, BoydSlowR}, storage of light\cite{M.F,Maynard}, nanoscale resolution\cite{AgarwalNano}, magnetometry \cite{ScullyMAG}, etc. The interference resulting from multiple path ways produces a narrow dip in the absorption spectra which typically are used in applications. The atomic coherence effects in single electron atoms like Na, $^{85}Rb$, and Cs have been extensively studies. Results for both coherent population trapping and coherent population oscillations are available \cite{Min,BoydSlow,AgarwalLocal,Miles,He}.  Compared to single-electron atoms, there are only much fewer studies for two-electron atoms. Mompart et al. have presented a study of CPT in two-electron atom with aligned spins \cite{Mompart}. In the investigation of a two electron system one needs to take in to account the Pauli exclusion principle to obtain the allowed transitions. Maynard et al \cite{Maynard} studied the transition $2$ $^{3}S_{1} \rightarrow 2$ $^{3}P_{1}$ in metastable He. This is a $ \Lambda $-system. They observed CPO between two levels involving only change of spin. Mompart et al \cite{Mompart} specifically studied the transitions in a two electron system where both ground and excited states had aligned spins (ortho system) so that the orbital part of the two electron system was antisymmetric. Using this they showed a very interesting possibility that an ortho systemو like the transition 4s4p to 4p4p in Ca, which is a $V$-system can show CPT. This is because of the Pauli principle that the $ V $-system becomes equivalent to a Lambda system.
In a recent experimental work Singh and Natarajan studied the intercombination line  $^{1}S_{0} \rightarrow $ $^{3}P_{1}$ in $^{174}Yb$ \cite{Singh}. They reported both CPO and CPT in such a system. It is to be noted that for the observation of CPO in two level atoms, we need strong dephasing \cite{BoydSlow}. An atomic beam has no dephasing, hence observation of CPO is quite remarkable. The ground state of $^{174}Yb$ has the configuration $ 6 s^{2} $  with the term symbol $ ^{1}S_{0} $ and the upper state is 6s6p with the term symbol $ ^{3}P_{1} $. They also reported both CPT and CPO by using a magnetic field. Note that in the ground state the two spins are antiparallel whereas in the excited state the two spins are parallel. Thus the situation is quite different from that considered by Mompart et al \cite{Mompart}. We thus need a theoretical model to understand the experimental results on the intercombination line, which is forbidden in LS coupling, of $^{174}Yb$. We also note that in the experimental study the pump and probe had comparable intensities and thus the probe unlike other experiments is not weak compared to the pump and the saturation of the transition by both pump and probe fields is expected to have a major effect. This must be accounted in any theoretical modelling.
The organization of this paper is as follows, in Sec II we describe the intercombination line as an effective four level system with the ground and excited states coupled by pump and probe fields which are orthogonally polarized. The two fields have comparable intensities. We derive the basic density matrix equations and present the expressions for the fluorescence. In Sec III we presents numerical results for fluorescence obtained from a Floquet analysis. We show results in the absence of the magnetic field and in presence of the magnetic field. Our numerical results confirm the behavior as observed in experiments. In Sec IV we present a number of analytical results when the magnetic field is zero. The analytical results help us understand the observed behavior of fluorescence.

\section{Model and Equations}

\begin{figure}
\centering
\resizebox{0.4\textwidth}{!}{%
  \includegraphics{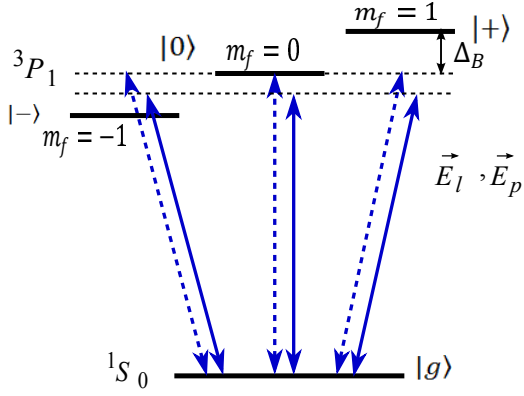}}
\caption{\small Two-electron four-level quantum system driven by two orthogonally polarized probe $\stackrel{\rightarrow}{E_{p}}$ (solid) and pump $\stackrel{\rightarrow}{E_{l}}$ (dashed) fields. As indicated in the text these couplings with the lasers are effective couplings.}
\label{model}
 \end{figure}

Consider an atomic model as shown in  Fig. \ref{model}. The ground state is $\left|g\right\rangle=$ $^{1}S_{0}\left(F=0, M_{f}=0\right)$ and the upper levels are $\left|-\right\rangle =$ $^{3}P_{1}\left( F=1, M_{f}=-1 \right) $, $\left|0\right\rangle = $ $ ^{3}P_{1}$ $\left(F=1, M_{f}= 0\right)$, ~and $~~\left|+ \right\rangle=$ $^{3}P_{1} \left( F=1, M_{f}=1\right)$. The transtion used is the $^{1}S_{0}\rightarrow $ $^{3}P_{1}$ at $556  nm$ intercombination line of the even isotop $^{174}Yb$ atom. The state $^{3}P_{1}$ is weakly mixed to transition$^{1}P_{1}$ which has two spins antiparallel. Thus \textit{effectively}, we can think that the level $^{3}P_{1}$ is coupled to the level $^{1}S_{0
}$ via the laser field. Two orthogonal linarly polarized fields, $\stackrel{\rightarrow}{E_{p}}$ as the probe field and $\stackrel{\rightarrow}{E_{l}}$ as the pump field, couple the ground level to the upper levels :
\begin{eqnarray}
\stackrel{\rightarrow}{E_{p}}& = &\varepsilon_{p}\left(sin\theta \hat{x}+cos\theta \hat{z}\right) e^{-i\omega_{p} t+ik_{p}y}+c.c, \nonumber \\
\stackrel{\rightarrow}{E_{l}}& = &\varepsilon_{l}\left(cos\theta \hat{x}-sin\theta \hat{z}\right) e^{-i\omega_{l} t+ik_{l}y}+c.c,\nonumber \\
\stackrel{\rightarrow}{E_{p}}& . &\stackrel{\rightarrow}{E_{l}}=0,
\label{E}
\end{eqnarray}
where $\varepsilon_{p}$ and $\varepsilon_{l}$ are the probe and pump fields amplitude, respectively. Moreover $\theta$ showes the angle between the polarization and direction of the propagation. The Rabi frequensies of the probe and pump fields can be defind as $2\Omega_{pi}=2 [\varepsilon_{p}\left(sin\theta \hat{x}+cos\theta \hat{z}\right)]\cdot \stackrel{\rightarrow}{d_{ig}}/ \hbar$ and $2\Omega_{li}=  2[\varepsilon_{l}\left(cos\theta \hat{x}-sin\theta \hat{z}\right)]$.  $\stackrel{\rightarrow}{d_{ig}}/ \hbar~ \left (i=+,0,-\right)$. The $\stackrel{\rightarrow}{d_{ig}}$ gives the dipole matrix element which can be written using Clebsch-Gordan coefficients as
\begin{equation}
\stackrel{\rightarrow}{d_{+g}}=\left|d\right|\hat{\epsilon}_{+},~~~~\stackrel{\rightarrow}{d_{0g}}=\left|d\right|\hat{z},
~~~~\stackrel{\rightarrow}{d_{-g}}=\left|d\right|\hat{\epsilon}_{-},
\label{d}
\end{equation}
where $\left|d\right|$ is the reduced dipole matrix element and $\hat{\epsilon}_{\pm}=(\hat{x}\pm i\hat{y}/\sqrt{2})$. It should be borne in mind that the parameter $d$ would include the coefficient of mixing with the level $ ^{1}P_{1}$.
By using Eq. \ref{d}, the Rabi frequencies can be written as
\begin{eqnarray}
\Omega_{p+}=\Omega_{p-}& = & \frac{A_{p}}{\sqrt{2}}sin\theta,~~~~\Omega_{p0}= A_{p}cos\theta,\nonumber\\
\Omega_{l+}=\Omega_{l-}& = & \frac{A_{l}}{\sqrt{2}}cos\theta,~~~~\Omega_{l0}= - A_{l} sin\theta,
\label{omegap}
\end{eqnarray}
where $A_{p}=|d| \varepsilon_{p}$ and $A_{l}= |d| \varepsilon_{l}$.
We use the same geometry as in \cite{Singh} i.e. atomic beam is moving in direction $z$, the lasers are propagating perpendicular to the direction of the atomic beam.
The Hamiltonian of the system interacting with two laser fields in the dipole and rotating wave approximations is given by:
\begin{equation}
H_{in}=-\hbar[(\sum_{j=+,0,-}(\Omega_{pi}e^{-i\omega_{p}t}+\Omega_{li}e^{-i\omega_{l}t})\left|j\right\rangle\left\langle g\right|)+c.c],
\end{equation}
where $ \omega_{p}$ and $\omega_{l}$ denote the frequencies of the applied fields. In the rotating frame, the density matrix equations ,which show the response of the medium to the field, are given by:
\begin{eqnarray}
\dot{\rho}_{++} & = &\rho_{g+}\left(\Omega_{p+}e^{- i\Delta t}+\Omega_{l+}\right)-i \rho_{+g}\left(\Omega_{p+}e^{ i\Delta t}+\Omega_{l+}\right) \nonumber \\
&& {} -\gamma_{+g} \rho_{++} \nonumber,\\
\dot{\rho}_{00} & = &i \rho_{g0}\left(\Omega_{p0}e^{- i\Delta t}+\Omega_{l0}\right)-i \rho_{0g}\left(\Omega_{p0}e^{ i\Delta t}+\Omega_{l0}\right) \nonumber \\
&& {} -\gamma_{0g} \rho_{00} \nonumber,\\
\dot{\rho}_{--} & = &i \rho_{g-}\left(\Omega_{p-}e^{- i\Delta t}+\Omega_{l-}\right)-i \rho_{-g}\left(\Omega_{p-}e^{ i\Delta t}+\Omega_{l-}\right) \nonumber \\
&& {} -\gamma_{-g} \rho_{--},\nonumber \\
\dot{\rho}_{g+} & = &-i \delta_{+} \rho_{g+}+i \left(\rho_{++}-\rho_{gg}\right)\left (\Omega_{p+}e^{ i\Delta t} + \Omega_{l+}\right) \nonumber \\
&& {} +i \rho_{0+}\left(\Omega_{p0}e^{ i\Delta t}+\Omega_{l0}\right)+i \rho_{-+}\left(\Omega_{p-}e^{ i\Delta t}+\Omega_{l-}\right)\nonumber \\
&& {} -\Gamma_{g+}  \rho_{g+},\nonumber \\
\dot{\rho}_{g0} & = &-i \delta_{0} \rho_{g0} + i \left (\rho_{00} - \rho_{gg}\right) \left(\Omega_{p0}e^{ i\Delta t} + \Omega_{l0}\right) \nonumber \\
&& {} + i\rho_{+0}\left(\Omega_{p+}e^{ i\Delta t}+\Omega_{l+}\right)+i \rho_{-0}\left(\Omega_{p-}e^{ i\Delta t}+\Omega_{l-}\right)\nonumber \\
&& {} -\Gamma_{g0}  \rho_{g0},\nonumber \\
\dot{\rho}_{g-} & = &-i\delta_{-}\rho_{g-}+i \left(\rho_{--}-\rho_{gg}\right)\left(\Omega_{p-}e^{ i\Delta t}+\Omega_{l-}\right) \nonumber \\
&& {} +i \rho_{+-} \left(\Omega_{p+}e^{ i\Delta t}+\Omega_{l+}\right)+i \rho_{0-}\left(\Omega_{p0}e^{ i\Delta t}+\Omega_{l0}\right) \nonumber\\
&& {} -\Gamma_{g-} \rho_{g-},\nonumber \\
\dot{\rho}_{0+} & = &-i\left(\delta_{+}-\delta_{0}\right)\rho_{0+}-i \rho_{0g}\left(\Omega_{p+}e^{ i\Delta t}+\Omega_{l+}\right) \nonumber \\
&& {} +i \rho_{g+} \left(\Omega_{p0}e^{ -i\Delta t}+\Omega_{l0}\right)-\Gamma_{0+} \rho_{g0},\nonumber \\
\dot{\rho}_{-+} & = &i\left(\delta_{-}-\delta_{+}\right)\rho_{-+}+i \rho_{g+}\left(\Omega_{p-}e^{ -i\Delta t}+\Omega_{l-}\right) \nonumber \\
&& {} -i\rho_{-g}\left(\Omega_{p+}e^{ i\Delta t}+\Omega_{l+}\right)-\Gamma_{-+} \rho_{-+},\nonumber \\
\dot{\rho}_{-0} & = &i\left(\delta_{-}-\delta_{0}\right)\rho_{-0}+i \rho_{g0}\left(\Omega_{p-}e^{ -i\Delta t}+\Omega_{l-}\right) \nonumber \\
&& {} -i\rho_{-g}\left(\Omega_{p0}e^{ i\Delta t}+\Omega_{l0}\right)-\Gamma_{-0} \rho_{-0},
\label{density matrix}
\end{eqnarray} 
where $\gamma_{ig}$ is the spontaneous decay rate from level $\left|i\right\rangle$ to level $\left|g\right\rangle$. The off diagonal element, $\rho_{ij}$ decays at the rate $\Gamma_{ij}=(\gamma_{ig}+\gamma_{jg})/2$. The parameter $\Delta=\omega_{p}-\omega_{l}$ denotes the probe field detuning with respect to the pump field. Also $\delta_{i}=\omega_{l}-\omega_{ig} (i=+,0,-)$ is the pump field detuning with the atomic resonance transition.
Because of the explicit time dependence in Eq. (\ref{density matrix}), the steady state has the Floquet expansion:
\begin{equation}
\rho_{ij}=\sum^{\infty}_{m=-\infty}\rho^{(m)}_{ij} e^{-i m \Delta t},
\label{Floquet}
\end{equation} 
where $m=0$ denotes the time-independent part of the density matrix elements.
 The positive frequency part of the electric field operator at detector is \cite{Agarwalbook}
\begin{eqnarray}
\stackrel{\rightarrow}{E}^{(+)}(\stackrel{\rightarrow}{r}, t)=-\frac{\omega^{2}_{l}}{c^{2}}\stackrel{\rightarrow}{n}\times[\stackrel{\rightarrow}{n}\times\stackrel{\rightarrow}{d}]
\frac{e^{ik_{l}r-i\omega_{l}t}}{r}e^{-ik_{l}\stackrel{\rightarrow}{n}.\stackrel{\rightarrow}{R}},
\label{positive E}
\end{eqnarray}
where the dipole moment operator is given by
\begin{equation}
\stackrel{\rightarrow}{d}=\sum_{i=+, 0, -}\stackrel{\rightarrow}{d}_{gi}\left|g\right\rangle \left\langle i\right|,
\label{do}
\end{equation}
and $\stackrel{\rightarrow}{r}$ denotes the point at which the fluorescence be measured. Atomic sample is located at $\stackrel{\rightarrow}{R}$. Moreover $\stackrel{\rightarrow}{n}=\frac{\stackrel{\rightarrow}{r}}{r}$ is the direction of the observation. According to the experiment set up, the fluorescence is collected in the direction perpendicular to the atomic beam and the laser beams i.e. in $x$ direction. The fluorescence is given by 
\begin{equation}
I=\left\langle \stackrel{\rightarrow}{E}^{(-)}(\stackrel{\rightarrow}{r}, t)~ . ~\stackrel{\rightarrow}{E}^{(+)}(\stackrel{\rightarrow}{r}, t)\right\rangle,
\label{correlation}
\end{equation}
where $\stackrel{\rightarrow}{E}^{(-)}$ denotes the negative frequency part of the electric field operator at the detector.
With $\stackrel{\rightarrow}{n}=\hat{x}$ in Eq. (8), the dc component of fluorescene can be written as
\begin{equation}
I=\frac{I_{0}}{2}[2\rho^{0}_{00}+\rho^{0}_{++}+\rho^{0}_{--}-\rho^{0}_{+-}-\rho^{0}_{-+}],
\label{i}
\end{equation}
where the constant $I_{0}$ will depend on $\omega_{l}$ and $|d|$. In our further consideration, we will refer to fluorescence in units of $I_{0}$. Note that the fluorescence has contributions from both excited state populations $ I_{p}=\frac{I_{0}}{2}(2\rho^{0}_{00}+\rho^{0}_{++}+\rho^{0}_{--}), $ as well as coherences $ I_{c}=-\frac{I_{0}}{2}(\rho^{0}_{+-}+\rho^{0}_{-+}) $.
\begin{figure}
\centering
\resizebox{0.46\textwidth}{!}{%
  \includegraphics{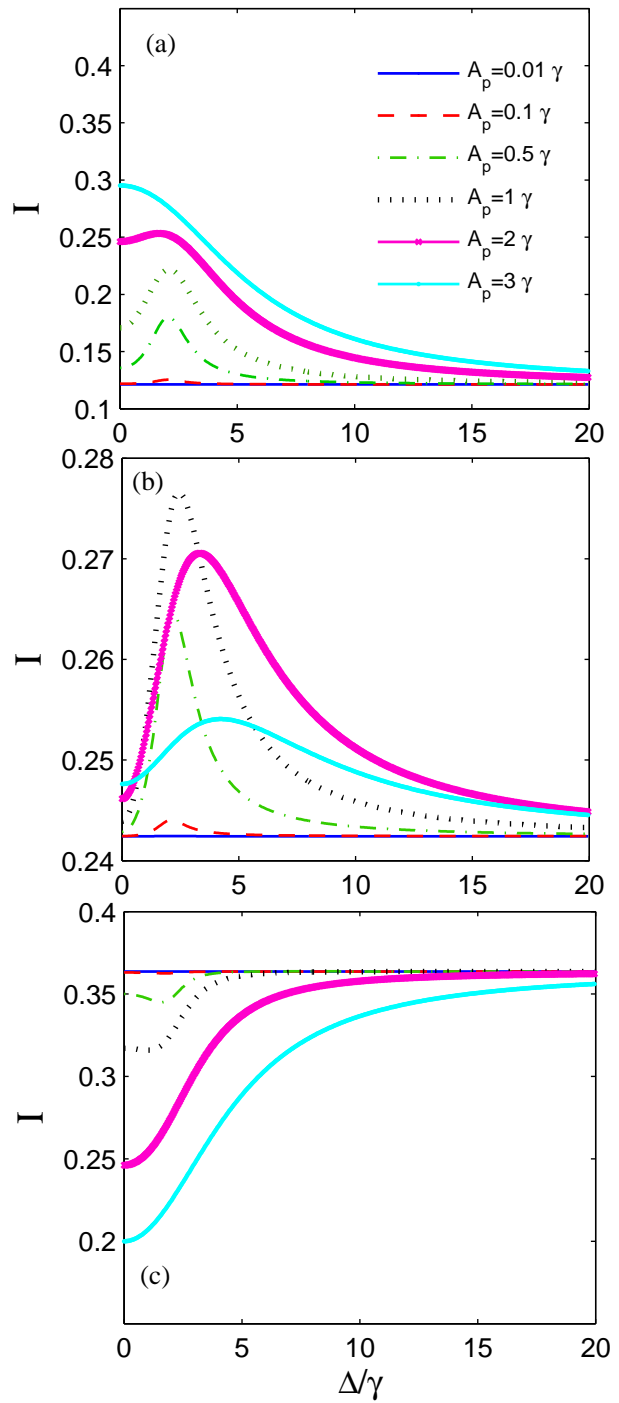}}
\caption{\small Fluorescence for different values of probe field. Selected parameters are $A_{l}=2\gamma$, $\theta=\pi/6$ (a),  $\theta=\pi/4$ (b) and $\theta=\pi/3$ (c).}
\label{pcf}
 \end{figure}
\begin{figure}
\centering
\resizebox{0.47\textwidth}{!}{%
  \includegraphics{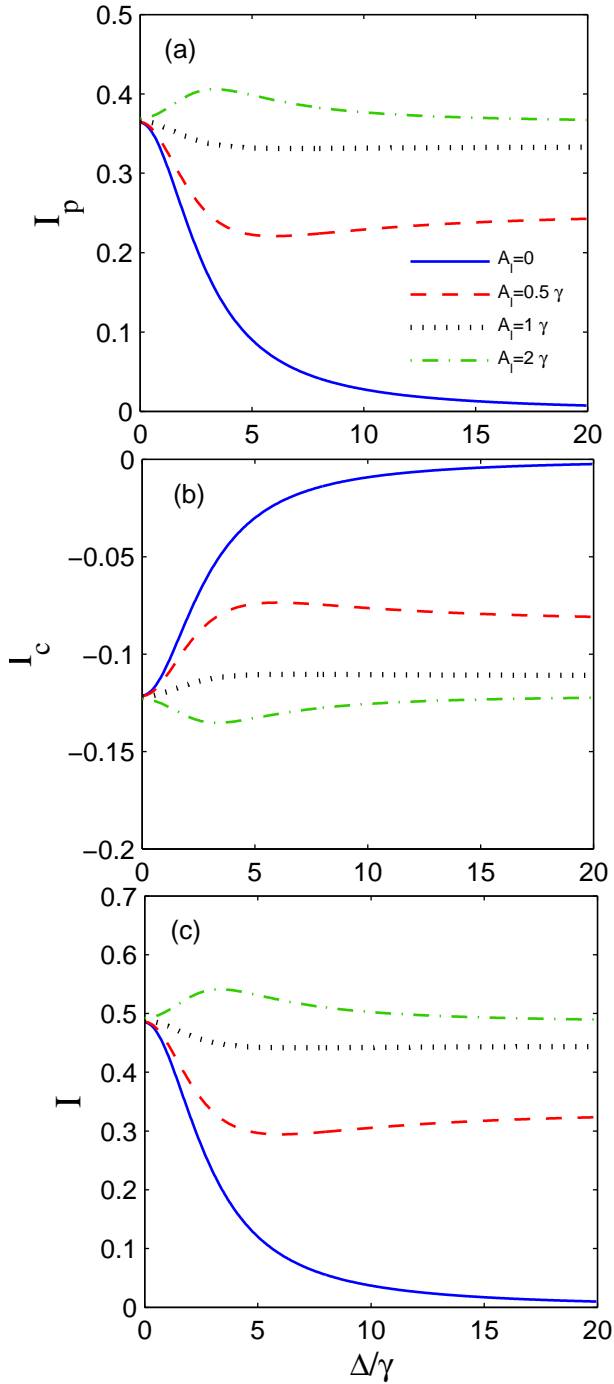}}
\caption{\small Population (a), coherence (b) terms and total magnitude of the fluorescence (c) for different values of pump field. Parameters used are $ \gamma_{+g}=\gamma_{-g}=\gamma_{0g}=\gamma$, $A_{p}=2\gamma$, $\delta_{\pm}=\delta_{0}=0$ and $\theta=\pi/4$.}
\label{pump}
 \end{figure}

\section{Numerical results}

The Eq. \ref{density matrix} for the density matrix elements are numerically solved using the Eq. \ref{Floquet}. Note that equations for $ \rho^{(m)}_{ij} $ get coupled to $ \rho^{(m \pm 1)}_{ij} $. The convergence of the truncation is tested for every set of parameters.
As in the experiment we choose pump on resonance i. e. we set $ \delta_{0}=0 $. The probe is scaned i. e. the detuning parameter $\Delta $ is varied. We scale all parameters in units of the natural line width $ \gamma=2 \pi \times 185 KHz$ for the $^{174}Yb$ intercombination line. Further $ \gamma_{\pm g}=\gamma_{0g}=\gamma $.
\begin{figure}
\centering
\resizebox{0.47\textwidth}{!}{%
  \includegraphics{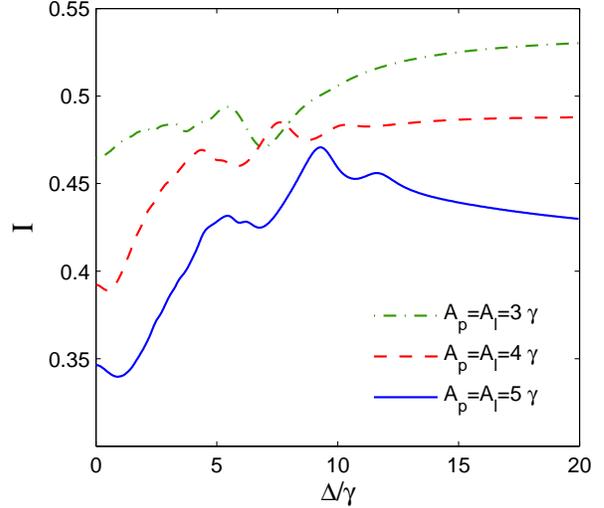}}
\caption{\small The fluorescence in presence of magnetic field. Parameters used are $\theta=\pi/4$ and $\Delta_{B}=4 \gamma$. The pump and probe are of equal intensities.}
\label{m}
 \end{figure}
 \begin{figure}
\centering
\resizebox{0.47\textwidth}{!}{%
  \includegraphics{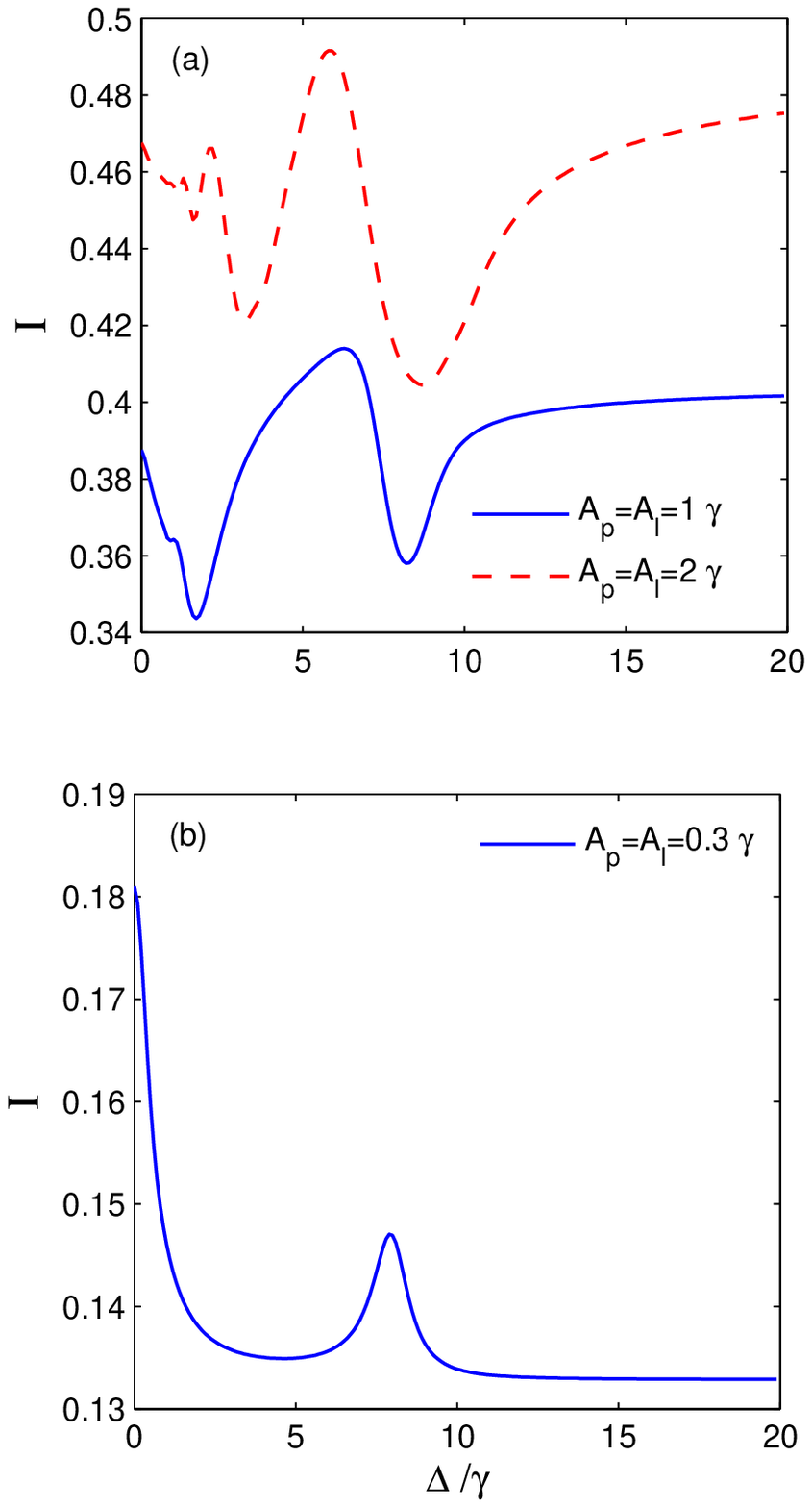}}
\caption{\small The fluorescence in presence of magnetic field. Parameters used are $\theta=\pi/4$ and $\Delta_{B}=8 \gamma$. The field strengths are as shown in the boxes.}
\label{m8}
 \end{figure}  
 
We first describe the results in the absence of the magnetic field. In figures, we show the fluorescence as a function of the probe detuning $\Delta\geq0 $ only as $ I(-\Delta)=I(+\Delta) $. Typically pump and probe experiments are done for a weak probe and strong pump. We first fix the Rabi frequency of the pump value at which the transition can be saturated ($ A_{l}=2\gamma $). We show fluorescence $I$ (in units of $I_{0}$) in Fig. \ref{pcf} for different values of the strength of the probe. Note that for $ \theta\neq\pi/4 $, the strength of the pump and probe for different transitions are different, for example $|g\rangle\rightarrow|0\rangle$ transition has probe (pump) Rabi frequency proportional to $ \cos\theta $ ($ \sin\theta $). We show the results for different polarization angles in Fig. \ref{pcf}. This figure shows pronounce differences between the cases $\theta=\pi/3$, $\theta=\pi/6$, and $\theta=\pi/4$. As the strength of the probe increases, a minimum at $\Delta=0$ starts appearing. For $ \theta=\pi/4$, the minimum at $ \Delta=0 $ is most pronounced when pump and probe strengths are comparable. The fluorescence behavior in Fig. \ref{pcf} is quite comparable to that reported in \cite{Singh}. The results of this figure clearly show that the experimentally observed dip at $\Delta=0$ is due to the saturation produced by both pump and probe fields. We also present the results for fluorescence when the probe saturates the atomic transition and the strength of the pump field is increased. As mentioned earlier, $I$ has contributions from populations and coherences. These contributions are shown separately in Figs. \ref{pump} a, b. The Fig. \ref{pump}c gives the interference contribution to $I$. For the direction of observation under consideration, the interference terms are destructive. It is shown that the observed dip in $ I $ at $\Delta=0$ is the result of the saturation of the transitions by both pump and probe fields. Further both populations and coherences contribute to the dip.\\
In the presence of magnetic field, the excited levels $ |\pm\rangle $ split by $ \Delta_B=\pm g \mu B $, where $g$ is the Lande g factor, $\mu$ is the bohr magneton and $B$ is the magnetic field. Now the CPT resonances and CPO resonances separate out. The CPO resonances still occur at $ \Delta=0 $. The CPT like resonances would occur at position determind by the magnetic field. In Fig. \ref{m}, we show the behavior of $I$ as a function of $ \Delta $ for $ \Delta_B=4 \gamma $ and for different values of the Rabi frequencies. The numerical results in Fig.  \ref{m} show the general trend seen in the experimental data. In Fig. \ref{m8} we present additional results for larger value of the magnetic field. We see a resonant structure at $ \Delta\sim\Delta_{B} $ for low Rabi frequencies. As the pump and probe saturate the atomic transition, a well defind interference minimum is seen at $ \Delta\sim\Delta_{B} $. This is in agreement with the experimental observation. We also see an additional minimum at $ \Delta\sim\Delta_{B}/2 $. This additional minimum can be interpreted as a subharmonic resonance. We note that subharmonic resonances were extensively studied in context of stimulated Raman scattering \cite{GSA,RT}. The subharmonic resonances arise from strong saturation by both pump and probe fields. However the numerical results do not quite yield the CPT resonance at $ \Delta=2\Delta_{B} $, as observed in the experiment. It should be borne in mind that the magnetic field can couple the level $ ^{3}P_{1} $ to the $ ^{1}P_{0} $ and $ ^{3}P_{2} $ levels \cite{coupling,magnetic} and this coupling is most likely the reason why our numerical results do not show the interference minimum at $ \Delta=2\Delta_{B} $. It may be note that the magnetic field coupling to such states would be like a phase perturbation and can give rise to interferences similar to collision induced effects \cite{BL}. 

\section{Analytical Results for Fluorescence in the absence of magnetic field}

Remarkably enough, the set of Eqs. \ref{density matrix} can be solved analytically in the absence of the magnetic field. Let us make a transformation to a new basis defined by 
\begin{eqnarray}
\left|\psi_{1}\right\rangle & = & (\frac{\left|+\right\rangle+\left|-\right\rangle}{\sqrt{2}})cos\theta  - \left|0\right\rangle sin \theta,  \nonumber \\
\left|\psi_{2}\right\rangle & = & (\frac{\left|+\right\rangle+\left|-\right\rangle}{\sqrt{2}})sin\theta + \left|0\right\rangle cos \theta, \nonumber \\
\left|\psi_{3}\right\rangle & = & \frac{\left|+\right\rangle-\left|-\right\rangle}{\sqrt{2}}.
\label{new basis}
\end{eqnarray}
\begin{figure}
\centering
\resizebox{0.3\textwidth}{!}{%
  \includegraphics{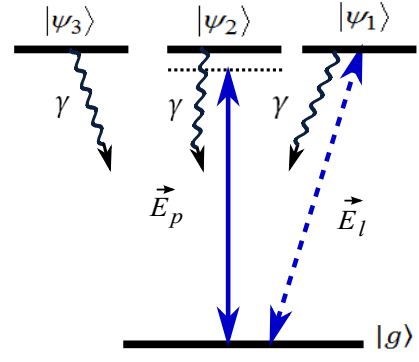}}
\caption{\small Schematic diagram of quantum system in new basis.}
\label{model2}
 \end{figure}  
The choice of these basis is determined by the polarization of the pump and probe fields. 
The states $ \left|\psi_{i}\right\rangle  (i=1, 2, 3)$ form an orthogonal set. The $ \left|\psi_{3}\right\rangle$ does not couple to either probe or pump fields. The level $ \left|\psi_{2}\right\rangle $ ($\left|\psi_{1}\right\rangle$) couples only to the level $ \left|g\right\rangle $ by the probe (pump) field. 
It can also be shown that all the decay rate $ \gamma_{\psi_{1}g},  \gamma_{\psi_{2}g} $ and $ \gamma_{\psi_{3}g}$ are equal to $ \gamma $. In the new basis the pump, probe fields and spontaneaus emission transitions are shown in Fig. \ref{model2}. We now rewrite Eqs. \ref{density matrix} in terms of the density matrix elements in the new basis  
\begin{equation}
\rho_{\alpha\beta}=\langle\psi_{\alpha}|\rho|\psi_{\beta}\rangle .
\label{new rho}
\end{equation}
Since the level $ \left|\psi_{3}\right\rangle $ is decoupled, it can be dropped from further consideration. The relevant density matrix equations are 
\begin{eqnarray}
\dot{\rho}_{\psi_{1}\psi_{1}} & = & i\rho_{g\psi_{1}}A_{l} -  i\rho_{\psi_{1}g} A_{l} - \gamma_{\psi_{1}g} \rho_{\psi_{1}\psi_{1}} \nonumber,\\
\dot{\rho}_{\psi_{2}\psi_{2}} & = & i\rho_{g\psi_{2}}A_{p} - i\rho_{\psi_{2}g} A_{p} - \gamma_{\psi_{2}g} \rho_{\psi_{2}\psi_{2}} \nonumber,\\
\dot{\rho}_{\psi_{1}\psi_{2}} & = & - i \Delta \rho_{\psi_{1}\psi_{2}} - i\rho_{\psi_{1}g} A_{p} + i\rho_{g\psi_{2}} A_{l}\nonumber \\
&& {} -\Gamma_{\psi_{1}\psi_{2}} \rho_{\psi_{1}\psi_{2}} \nonumber,\\
\dot{\rho}_{g\psi_{1}} & = &i(\rho_{\psi_{1}\psi_{1}} - \rho_{g g}) A_{l}+i\rho_{\psi_{2}\psi_{1}} A_{p}-\Gamma_{g\psi_{1}} \rho_{g\psi_{2}} \nonumber,\\
\dot{\rho}_{g\psi_{2}} & = &-i \Delta \rho_{g\psi_{2}}+i(\rho_{\psi_{2}\psi_{2}}-\rho_{g g}) A_{p}+i\rho_{\psi_{1}\psi_{2}} A_{l} \nonumber \\
&& {} -\Gamma_{g\psi_{2}} \rho_{g\psi_{2}}.
\label{new density matrix}
\end{eqnarray} 
In driving Eq. \ref{new density matrix}, we use a different rotating frame so that no explicit time dependence appears in Eq. \ref{new density matrix}. The state $ \left|\psi_{1}\right\rangle $ is rotated with the pump frequency and the state $ \left|\psi_{2}\right\rangle $ is rotated with the probe frequency. 
The fluorescence in new basis is
\begin{equation}
I=\frac{I_{0}}{2}[\rho_{\psi_{1}\psi_{1}}(1-\cos 2\theta)+\rho_{\psi_{2}\psi_{2}}(1+\cos 2\theta)].
\label{new I}
\end{equation}
The Eqs. \ref{new density matrix} are for $V$-system and one can solve for arbitary strenghs of the pump and probe fields. The full solutions for $ \rho_{\psi_{1}\psi_{1}} $ and $\rho_{\psi_{2}\psi_{2}}$ are
\begin{eqnarray}
\rho_{\psi_{1}\psi_{1}}& = &\frac{4A^{2}_{l}(N_{1}+N_{2})}{M},\nonumber\\
\rho_{\psi_{2}\psi_{2}}& = &\frac{4A^{2}_{p}(N_{3}+N_{4})}{M},
\label{T}
\end{eqnarray}
where 
\begin{eqnarray}
N_{1}& = &4 A^{4}_{p}+4 A^{4}_{l}+\gamma^{4}+5 \gamma^{2}\Delta^{2}+4 \Delta^{4},\nonumber\\
N_{2}& = &4 A^{2}_{l}(\gamma^{2}-2 \Delta^{2})+4A^{2}_{p}(2A^{2}_{l}+\gamma^{2}+4\Delta^{2}),\nonumber\\
N_{3}& = &4 A^{4}_{p}+4 A^{4}_{l}+4 A^{2}_{p}(2A^{2}_{l}+\gamma^{2}),\nonumber\\
N_{4}& = &\gamma^{2}(\gamma^{2}+\Delta^{2})+4 A^{2}_{l}(\gamma^{2}+\Delta^{2}),\nonumber\\
M &=& 32 A^{6}_{p}+4 A^{4}_{p}(24 A^{2}_{l}+9\gamma^{2}+4\Delta^{2})+(8 A^{2}_{l}+\gamma^{2})\nonumber\\
&&{} (4 A^{4}_{l}+\gamma^{4}+5\gamma^{2}\Delta^{2}+4 \Delta^{4}+4 A^{2}_{l}(\gamma^{2}-2\Delta^{2}))\nonumber\\
&&{}  +12 A^{2}_{p}(8 A^{4}_{l}+\gamma^{4}+2\gamma^{2}\Delta^{2}+6 A^{2}_{l}(\gamma^{2}+2 \Delta^{2})).\nonumber
\end{eqnarray}
Using analytical results Eqs. \ref{II} and \ref{T} we have reproduced the numerical results of Figs. \ref{pcf} and \ref{pump}. Now the analytical result is used to find the strength of the pump and probe for which the interference minimum would appear. 
For $\theta=\pi /4$, and $ \Delta $ in the neighborhood of zero, $ I $ becomes 
\begin{eqnarray}
I &=& \frac{4( A^{2}_{p}+ A^{2}_{l})}{8 A^{2}_{p}+8 A^{2}_{l}+\gamma^{2}}\nonumber\\
&& {} - \frac{16 A^{2}_{p}(2 A^{2}_{p}-6 A^{2}_{l}+\gamma^{2})\Delta^{2}}{(2 A^{2}_{p}+2 A^{2}_{l}+\gamma^{2})(8 A^{2}_{p}+8 A^{2}_{l}+\gamma^{2})^{2}}.
\label{II}
\end{eqnarray}
Clearly for no pump $ B=0 $, $ I $ has a peak at $ \Delta=0 $, as can be seen from inspection or from $ \frac{\partial^{2}I}{\partial\Delta^{2}}<0 $. The peak crosses over to a dip at a pump power given by 
\begin{equation}
A^{2}_{l}>\frac{2 A^{2}_{p}+\gamma^{2}}{6}
\label{final}
\end{equation}
Our numerical results in Figs. \ref{pump} and \ref{pcf} are in conformity with the analytical result, Eq. \ref{final}.

\section{Conclusions}  
We have presented theoretical modelling of the experiments on coherent population oscillations and coherent population trapping on the intercombination line of $^{174}Yb$. The transition involves a change of the spin and thus can not be interpreted in terms of an effective Lambda system which was suggested in \cite{Singh} using the theoretical framework of \cite{Mompart}. The reported experiments are done in the regime where both pump and probe fields can saturate the transition. We have shown by both numerical and analytical calculations the appearance of the interference minimum as both pump and probe start saturating the transition. We present an analytical result for the threshold probe power when the interference minimum can appear. We also present detailed study of the appearance of the interference minimum when magnetic fields are applied. Our studies show the newer possibility of the appearance of the subharmonic resonances in suitably chosen range of the pump and probe powers. The magnetic fields not only create Zeeman splittings but in addition make the system open because of the couplings to other levels. Such a coupling can give rise to additional resonances in a way similar to the dephasing induced resonances.


\begin{thebibliography}{99}                                                                                               


\bibitem {Arimondo}E. Arimondo, in \textit{Progress in Optics}, vol 35,
Ed. E. Wolf (North Holland, Amsterdam, 1996) pp. 258-354; G. Alzetta, A. Gozzini, L. Moi, and G. Orriols, Nuovo Cimento B, \textbf{36}, 5 (1976); E. Arimondo and G. Orriols, Lett. Nuovo Cimento  \textbf{17}, 333 (1976); H. R. Gray, R. M. Whitley, and C. R. Stroud, Opt. Lett. \textbf{3}, 6 (1978). 


\bibitem {BoydCPO}L. W. Hillman, R. W. Boyd, J. Krasinski, and C. R. Stroud, Opt. Commun, \textbf{45}, 36 (1983).

\bibitem {HarrisEIT}K.J. Boller, A. Imamoglu, S. E. Harris, \prl \textbf{66} 2593 (1991); S. E. Harris, Phys. Today, \textbf{50}, 36 (1997).

\bibitem{AgarwalLWI} O. Kocharovskaya and Ya. I. Khanin, Pisma Zh. Eksp. Teor. Fiz. \textbf{48}, 581 (1988)
 (JETP Lett. \textbf{48}, 630 (1988)); S. E. Harris, \prl\textbf{62}, 1033 (1989); 
 M. O. Scully, S.-Y. Zhu, and A. Gavrielides, \prl \textbf{62}, 2813 (1989);
  E. S. Fry, X. Li, D. Nikonov, G. G. Padmabandu, M. O. Scully, A. V. Smith, F. K. Tittel, C. Wang, S. R. Wilkinson, and S.-Y. Zhu, \prl \textbf{70}, 3235 (1993).
 
\bibitem{ScullyRef} M.O. Scully, \prl \textbf{67}, 1855 (1991). 

\bibitem{Slow} L.V. Hau, S. E. Harris, Z. Dutton, C.H. Behroozi, Nature  \textbf{397}, 594 (1999).

\bibitem{SlowA} S. E Schwarz and T. Y. Tan, Appl. Phys. Lett. \textbf{10}, 4 (1967).
  
\bibitem{BoydSlowR} G. S. Agarwal and T. N. Dey, Laser  Photon. Rev. \textbf{3}, 287 (2009);  R. W. Boyd, D. J. Gauthier in \textit{Progress in Optics}, vol 43,
Ed. E. Wolf (North Holland, Amsterdam, 2002) pp. 497-530.

\bibitem{M.F} M. Fleischhauer and M. D. Lukin, \prl\textbf{84}, 5094 (2000).

\bibitem{Maynard} M. A. Maynard, F. Bretenaker, and F. Goldfarb, \pra  \textbf{90}, 061801 (2014).

\bibitem{AgarwalNano} K. T. Kapale, and G. S. Agarwal, Opt. Lett, \textbf{35}, 2792 (2010). 

\bibitem{ScullyMAG} M. O. Scully, and M. Fleischhauer, \prl\textbf{69}, 1360 (1992).

\bibitem{Min} M, Xiao, Y. Q. Li, S. Z. Jin, and J. G. Banacloche, \prl\textbf{74}, 666 (1995); F. S. Cataliotti, C. Fort, T. W. Hansch, M. Inguscio, and M. Prevedelli, \pra\textbf{56}, 2221 (1997).

\bibitem{BoydSlow} M. S. Bigelow, N. N. Lepeshkin, and R. W. Boyd, \prl\textbf{90}, 113903 (2003).

\bibitem{AgarwalLocal} G. S. Agarwal, and K. T. Kapale, J. Phys. B \textbf{39}, 3437 (2006). 

\bibitem{Miles} J. A. Miles, D. Das, Z. J. Simmons, and D. D. Yavuz, \pra\textbf{92}, 033838 (2015).

\bibitem{He} T. Laupretre, S. Kumar, P. Berger, R. Faoro, R. Ghosh, F. Bretenaker, and F. Goldfarb, \pra\textbf{85}, 051805 (2012).

\bibitem{Mompart} J. Mompart, R. Corbalan, and L. Roso, \prl\textbf{88}, 88023603 (2001).

\bibitem{Singh} A. K. Singh and V. Natarajan, New J. Phys. \textbf{17}, 033044 (2001).

\bibitem{Agarwalbook} G. S. Agarwal,  \textit{Quantum Optics}, (Cambridge University Press, New York, 2013).

\bibitem{GSA} G. S. Agarwal, Opt. Lett. \textbf{13}, 482 (1988).

\bibitem{RT} R. Trebino and L. A. Rahn, Opt. Lett. \textbf{12}, 912 (1987).

\bibitem{coupling} L. J. Curtis and D. G. Ellis, J. Phys. B: At. Mol. Opt. Phys. \textbf{29}, 645 (1996).

\bibitem{magnetic} A.V. Taichenachev and V. I. Yudin, \prl \textbf{96}, 083001 (2006).

\bibitem{BL}  N. Bloembergen, H. Lotem, R.T. Lynch, Indian J. Pure and Appl. Phys. \textbf{16}, 151 (1978). 
 
\end{thebibliography}
\end{document}